\documentclass[a4paper,fleqn]{cas-dc}
\usepackage[numbers,longnamesfirst]{natbib}
\usepackage{amsmath}
\usepackage{float}
\usepackage{physics}
\usepackage{lipsum}
\usepackage{graphicx}
\usepackage{ulem}
\usepackage{footnote}
\usepackage{xcolor}
\usepackage{lineno}
\def\tsc#1{\csdef{#1}{\textsc{\lowercase{#1}}\xspace}}
\tsc{WGM}
\tsc{QE}
\tsc{EP}
\tsc{PMS}
\tsc{BEC}
\tsc{DE}
\begin{document}
\let\WriteBookmarks\relax
\def\floatpagepagefraction{1}
\def\textpagefraction{.001}
\shorttitle{Femtosecond Filament Coupled with Structured Light for Free Space Optical Communication}
\shortauthors{S.~B.~Ali~Reza, T.~Wang et~al.}
\title [mode = title]{Femtosecond Filament Coupled with Structured Light for Free Space Optical Communication}                      
\author[1]{\color{black} Saad Bin Ali Reza}
\fnmark[1]
\author[2]{\color{black} Tianhong Wang}
\fnmark[1]
\author[2]{\color{black} Finn Buldt}
\author[2]{\color{black} Pascal Bass\`ene}
\cormark[1]
\ead{bassep@rpi.edu}
\cortext[cor1]{Corresponding author}
\fntext[fn1]{These authors contributed equally to this work.}
\affiliation[1]{organization={Electrical, Computer, and Systems Engineering Department, Rensselaer Polytechnic Institute},
addressline={110 8th street}, city={Troy}, postcode={12180}, state={NY}, country={USA}}
\affiliation[2]{
organization={Department of Physics, Applied Physics and Astronomy, Rensselaer Polytechnic Institute},
addressline={110 8th street}, city={Troy}, postcode={12180}, state={NY}, country={USA}}
\author[2]{\color{black} Moussa N'Gom}
\begin{abstract}
The generation of a laser filament through clouds produces a shockwave which displaces the small water droplets from the air to create a quasi-transparent channel through which, information can be transmitted. We present a robust method that utilizes the channel for free-space optical communication using structured light beams coupled with a femtosecond filament. Vortex beams based on their spatial profiles and non-diffractive characteristics are perfectly suited to propagate around the filament. We have also developed a 4-bit communication channel using these structured light beam. We introduce a method dubbed segmented space division multiplexing. Our system demonstrates resilience to noise and is unaffected by the filament. This method can improve the scalability, robustness, and capacity of the free optical channel.
\end{abstract}
\maketitle
\section{\rm Introduction}
Light based communication between orbiting satellites and Earth's surface offers the prospect of significantly increasing space to ground data rates and constitutes a key element in future for secure worldwide quantum communication networks\cite{Bell:1880, Khalighi:2014, Israel:2017}. 
Free space optical links between Earth and space referred to as Free Space Optical  Communication (FSO) faces a persistent nemesis in the form of atmospheric clouds. The randomness in size and position of water droplets, which make up a rain cloud leads to substantial scattering of the optical energy and quickly scramble the signal encoded in thin laser beams. The amplitude fluctuation and wave-front distortion caused by atmospheric turbulence can additionally severely degrade the coupling efficiency and increase the bit-error-rate.  This barrier is currently surmounted by increasing the number of networked ground station, a very complex and expensive solution.
\\
Early attempts to clear the sky from fog and clouds with high power CW CO$_2$ laser mainly for increasing visibility have been realized. However, a very high energy is required to vaporize and shatter water drops (typically 10 kW.cm$^2$ and 10$-$1000 MW.cm$^2$ by pulse \cite{Kwok:1988, Pustovalov:1992}). Other methods use an aperture averaging technique to attenuate the amplitude fluctuation utilizing adaptive optics (AO) methods to compensate for the wave-front distortion caused by atmospheric turbulence \cite{Lee:2016, Leonhard:2016}. The emergence of femtosecond (fs) terawatt class lasers is an opportunity to reconsider FSO through dense clouds or fog with a fundamentally different approach: nonlinear propagation in the atmosphere and laser filamentation \cite{Khalighi:2014, Kasparian:2003, Schimmel:2018}.  
Compared to radio frequency (RF) communication, FSO operates at higher frequencies with wide-open bandwidth, resulting in significantly higher capacity communication links. 
One promising scheme to tackle signal losses in FSO, is to couple a laser filament together with a light beam carrying the signal. The filament displaces water droplets in its immediate vicinity to create a quasi-transparent channel within which the signal beam can travel unobstructed \cite{Schimmel:2018}. 
\\
Filamentation is a phenomenon describing a long thin plasma channel (the filament), produced by the balance between the optical Kerr effect and plasma defocusing \cite{Kelley:1965, Askaryan:1974, Braun:1995, Chin:2005, Chin:2012, Couairon:2007}. 
Laser filaments are self-sustained light of typically dozen micrometers diameter and up to hundreds of meters in length, widely extending the traditional linear diffraction limit \cite{Durand:2013, Polynkin:2008, Scheller:2014, Milles:2015, Hong:2016}.  
Properties of filaments such as their stability under air turbulence \cite{Chin:2002, Houard:2008} and their interaction with water droplets  have been studied \cite{Mechain:2005}. 
\\ 
The creation of the filament is accompanied by an expanding shock-wave that creates a cylindrical quasi-transparent channel around the filament \cite{Schimmel:2018, Yan:2020}. 
Schimmel, \textit{et al.} used a high repetition rate laser to demonstrate that the cleared channel can become  a stable millimeter-sized  waveguide virtually clear of obstructions  \cite{Schimmel:2018}.
\\
Gaussian beams, coupled to a filament have been used as signal carriers \cite{Schimmel:2018}. However, the Gaussian signal can be affected by the filament.  The dispersive effects of the filament can cause the information signal to be degraded. This, in addition to scattering from the remaining water droplets, are a serious barrier for the implementation of this method.

Here, we propose to embed the filament within a donut shaped light beam that can be used as the information carrier.  The signal carrying light are structured light with non-diffractive properties that  we generate dynamically using a spatial light modulator (SLM). 
The SLM allows us to create and control higher order Laguerre-Gauss (LG) and Bessel-Gauss (BG) beams, which are annular vortex beams possessing orbital angular momentum (OAM). 
Unlike Gaussian beams, the transverse intensity profile of these vortex beams have minimal diffraction as they propagates over a finite distance.  The SLM further allows us to control the diameter of the annular beam to fit within the cleared channel.
We then demonstrate a 4-bit Segmented Space Division Multiplexing (SSDM) scheme using a higher order LG beam. The principle behind the SSDM technique is to spatially divide the annular beam into segments, which can be modulated individually. Each segment of the donut shaped beam is generated using its own modified holographic mask.
%
\section{\rm Experimental Setup}
\begin{figure*}[th]
\centering
\includegraphics[width=.95\textwidth]{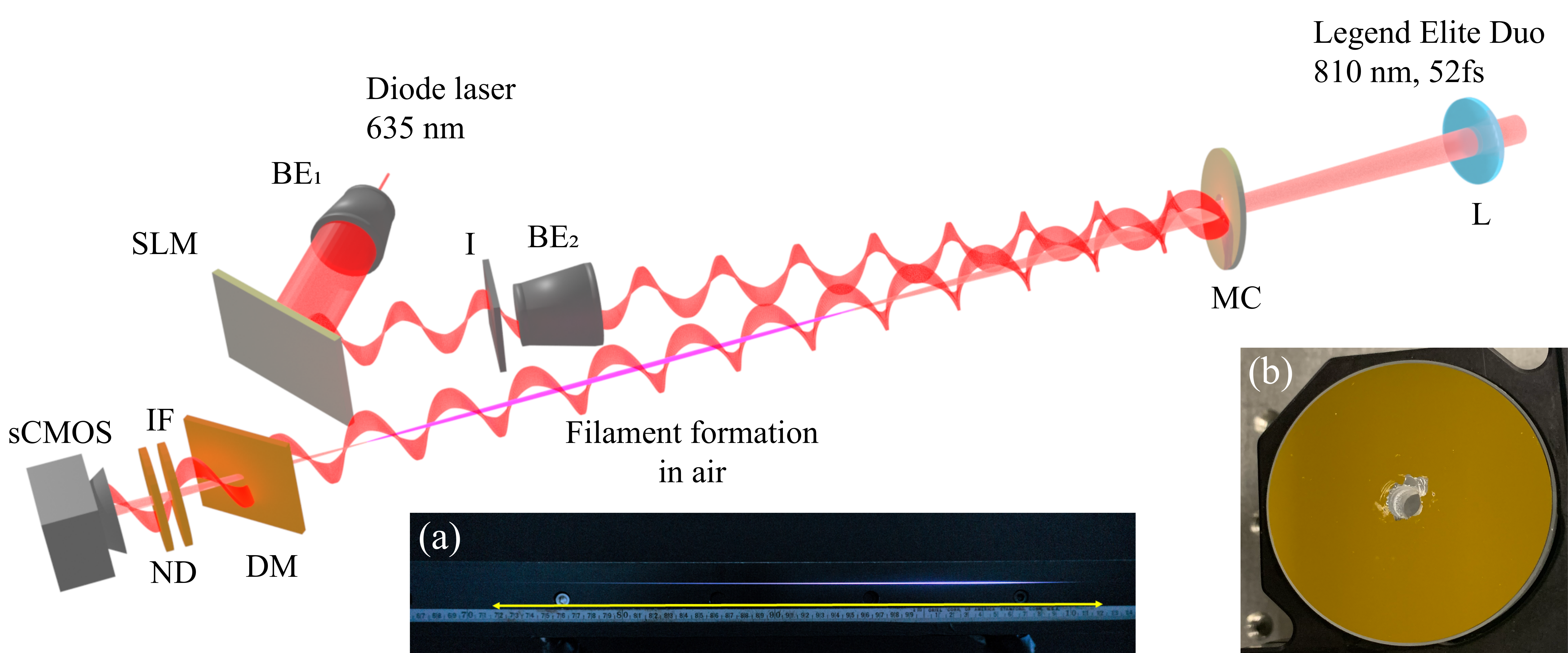}
\caption{Schematic of the experimental setup. L: lens $f = 2$ m, BE: beam expander, SLM: spatial light modulator, I: Iris, DM: dichroic mirror, IF: interference filters, ND: neutral density filters, MC: mirror coupler. Structured light beam generated by the SLM is coupled with the filament through the MC. After filtering out the femtosecond pulse with a DM, IF, and ND, the structured light is imaged by an sCMOS camera. Inset (a): Side view picture of the filament (glowing line) in air with a yellow line as a guide. Inset (b): Image of the in-house developed MC.}
\label{fig:setup}
\end{figure*}

An illustration of our experimental setup is shown in Figure \ref{fig:setup}. The femtosecond laser  is a \textit{Coherent Legend Elite Duo} with center wavelength at $\lambda$ = 810 nm.  Its  repetition rate is operated at 3 kHz with a measured pulse width of 52 fs, with  pulse energy at 1.67 mJ. 
The femtosecond beam is focused through air by a 2 m focal length anti-reflection (AR) coated lens. The side view of the resulting filament is imaged using a {\it Sony $\alpha$ a5000} camera and is shown in Figure \ref{fig:setup}$-$(a), where the white glowing line is the filament. A ruler is placed behind the filament as a reference.\\ 
The light carrying signal is  a continuous wave (CW), single longitudinal mode diode laser with wavelength $\lambda$ = 635 nm. The signal beam is collimated and expanded using a beam expander (BE$_1$). The diameter of the beam is chosen so that it covers the entire face of the SLM. The light modulator is a \textit{Santec SLM-200} phase-only SLM. 
\\
We display various holographic masks on the SLM to generate any desired structured light beams \cite{Arrizon:2007}. 
 An iris (I) is used to block out the undesired part of the signal. The remaining light is expanded and collimated using a second beam expander BE$_2$ before the designed mirror coupler (MC). 
The in-house developed MC is pictured in Figure \ref{fig:setup}--(b). It is a 2" (50.8 mm) gold mirror in which, we have drilled a hole approximately 6.3 mm in diameter.
\\
The MC is placed at the focal point of lens (L) so that the entire femtosecond beam  passes through the hole. The advantage of using a MC as opposed to a traditional beam combiner is that the femtosecond beam in its entirety is utilized to produce the filament. In general this will always result in a filament with higher plasma density \cite{Theberge:2006}. 
 A combination of dichroic mirror (DM), interference filters (IF), and neutral density filters (ND) are placed before the detectors to reduce the intensity of the femtosecond beam. The coupled lasers are imaged using  a Thorlabs sCMOS camera.
\section{\rm Results and Discussion}
The filaments we produce are a little less than a meter in length, limited by the lab environment. 
The critical power ($\mathcal{P}_{cr}$) for filamentation to occur is governed by \cite{Couairon:2007}: 
\begin{equation}
 \mathcal{P}_{cr}=\frac{\lambda}{8\pi C n_0 n_2}
\label{equ:p_crit}
\end{equation}
where $\lambda$ is the wavelength of the laser, $C$ is the numerical factor defined by the beam profile, $n_0$ is the refractive index of the medium (air), and $n_2$ is the nonlinear refractive index due to the optical Kerr effect.
\\ 
Once we produce the desired length filament, we proceed to embed in the different vortex beam different beams; eg. LG and BG beams. We also discuss the possibility of using an array of Gaussian beams and zeroth order BG beam to discuss the effect of the filament on Gaussian like beams. 
\subsection{\small\rm\bf LG Beam and Segmented Space Division Multiplexing}
\begin{figure}[h]
\centering
\includegraphics[width=\linewidth]{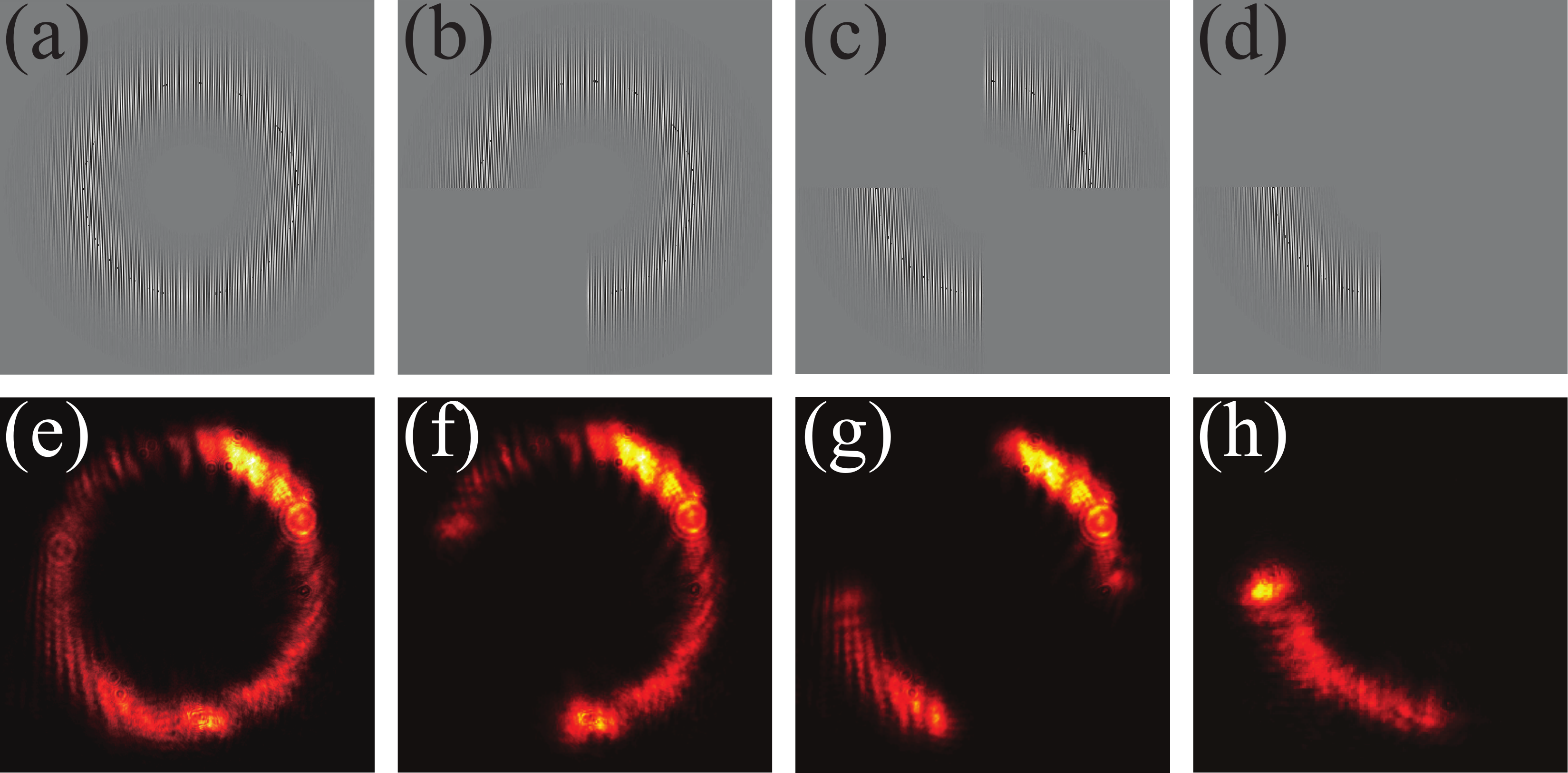}
\caption{Holographic masks displayed on the SLM to generate the intensity profile of the segmented LG beams. 
         In the top row (a--d) are shown the masks used to generate respectively the donut LG beam (e) and 
         the 1/4 (f), 2/4 (g), and 3/4 (h) segmented LG beams.}
\label{fig:mask}
\end{figure}

The phase singularity at the center of a vortex beam along its propagation axis, ensures no light is directly interacting with the filament. Also, the spatial profile and the immunity to diffractive spreading should allow it to effectively work in the quasi-transparent channel cleared by a filament. We produce a Laguerre-Gauss (LG) beam by illuminating the SLM which is displaying the corresponding phase mask \cite{Arrizon:2007}. 
\begin{figure}[h]
\centering
\includegraphics[width=0.95\linewidth]{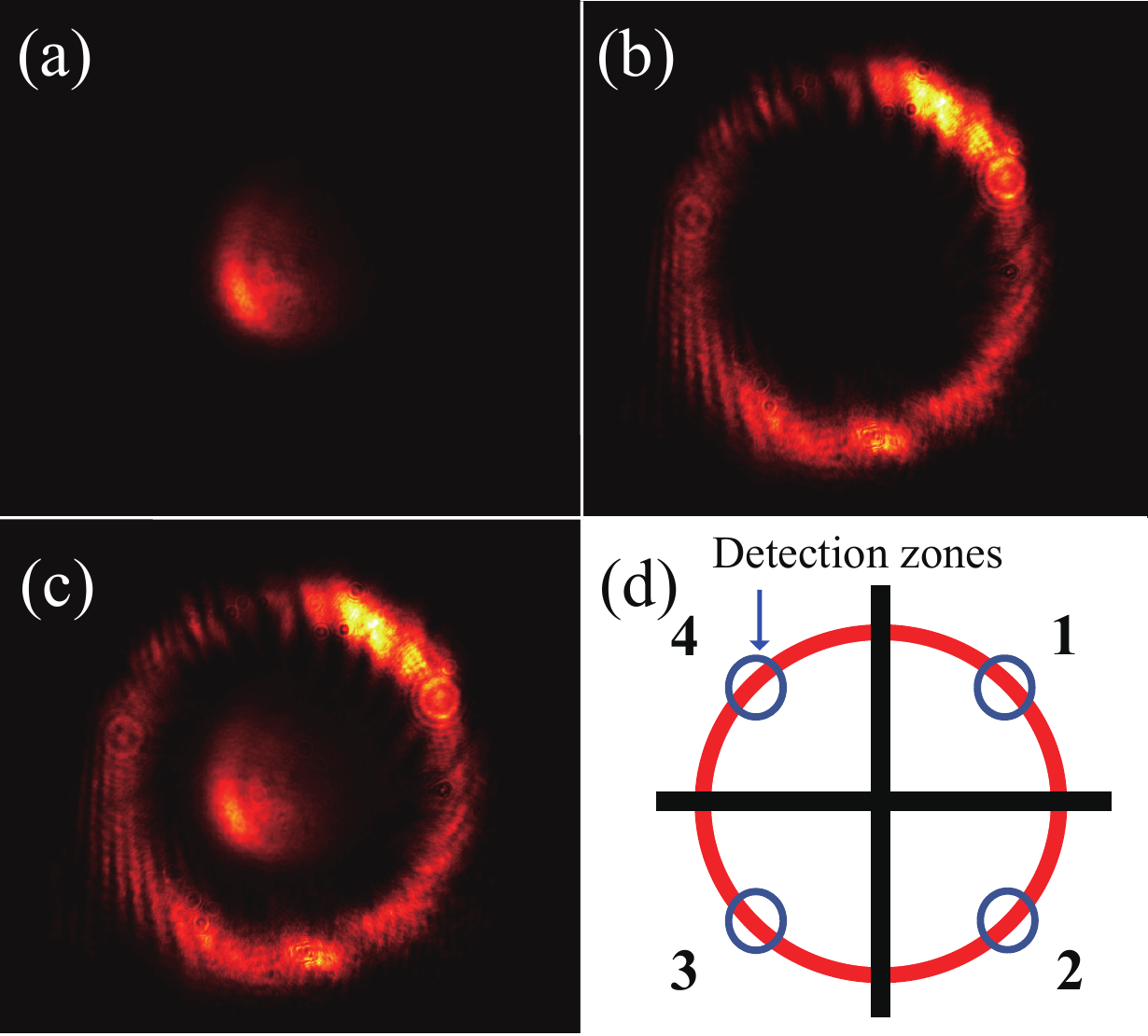}
\caption{(a) The femtosecond beam after the filament and filtering. (b) The LG ($\ell=12$) beam. (c) The coupled beams. (d) Segmented space division multiplexing mechanism. The ring is divided into 4 quadrants representing 4 bits. Each quadrant is controlled separately by the SLM.}
\label{fig:LG_nomod}
\end{figure}\\
The inner diameter of the LG beam is related to its topological charge ($\ell$) and the beam waist ($w\sim1$ cm) by $w\sqrt{2\ell}$. To correctly reflect the donut beam, the minimum ratio between the inner diameter of the LG beam and the diameter of the hole in the in-house developed MC (Figures \ref{fig:setup}$-$(b)) equals 7. We use an LG beam with $\ell =12$ which does not interact with the filament.\\
Figure \ref{fig:LG_nomod} depicts how the filament is embedded into the LG beam. The femtosecond beam shown in Figure \ref{fig:LG_nomod}--(a),  is attenuated by several orders of magnitude after the filament to avoid damaging the detectors. 
The LG signal and the fully embedded filament are shown in Figures \ref{fig:LG_nomod}--(b) and \ref{fig:LG_nomod}--(c). We vary the inner diameter of the annular beam between 7.6$-$10 mm, using beam expender (BE$_2$). This allows an extra flexibility to ensure that the annular beam fits well within the cleared channel.\\
The basic principal of SSDM is to divide the LG beam into 4 quadrants, each representing a bit. A visual representation of this is shown in Figure \ref{fig:LG_nomod}--(d). Since the LG beam is fully controlled by the phase mask on the SLM, each quadrant can be controlled individually by manipulating the phase mask \cite{Arrizon:2007}. The middle of each quadrant is selected as the detection zone to maximize the efficiency of the scheme. \\ 
The SSDM signals (Figure \ref{fig:mask}(e)$-$\ref{fig:mask}(h)) are produced by specific phase masks displayed on the SLM screen. The phase masks (Figure \ref{fig:mask}(a)$-$\ref{fig:mask}(d)) that generate the segmented beams are modified versions of the LG phase mask.
 A binary function is added on top of each super pixel or quadrant of the phase mask such that a single LG beam is divided into four independent segments. This enables the transmission of four multiplexed channels over one carrier. 
   \begin{figure}[h]
   \centering
   \includegraphics[width=.5\textwidth]{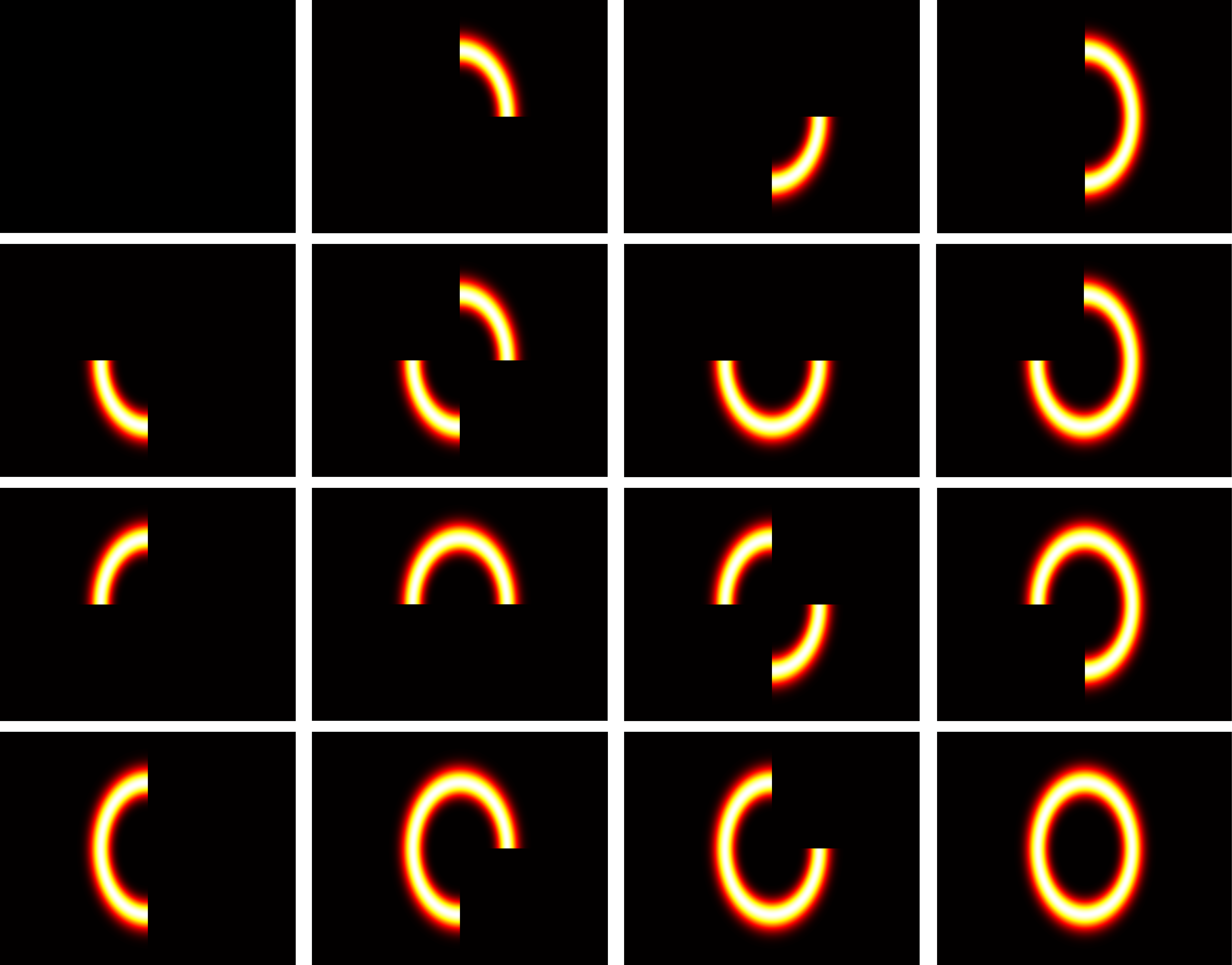}
   \caption{Simulation of 4-segmented bits by modulating an LG beam. Here are represented all the combinations of the 4-bit segmented system from 0000 (top left) to  1111 (bottom right).}
   \label{fig:bits}
  \end{figure}
\begin{figure}[h!]
\centering
\includegraphics[width=0.8\linewidth]{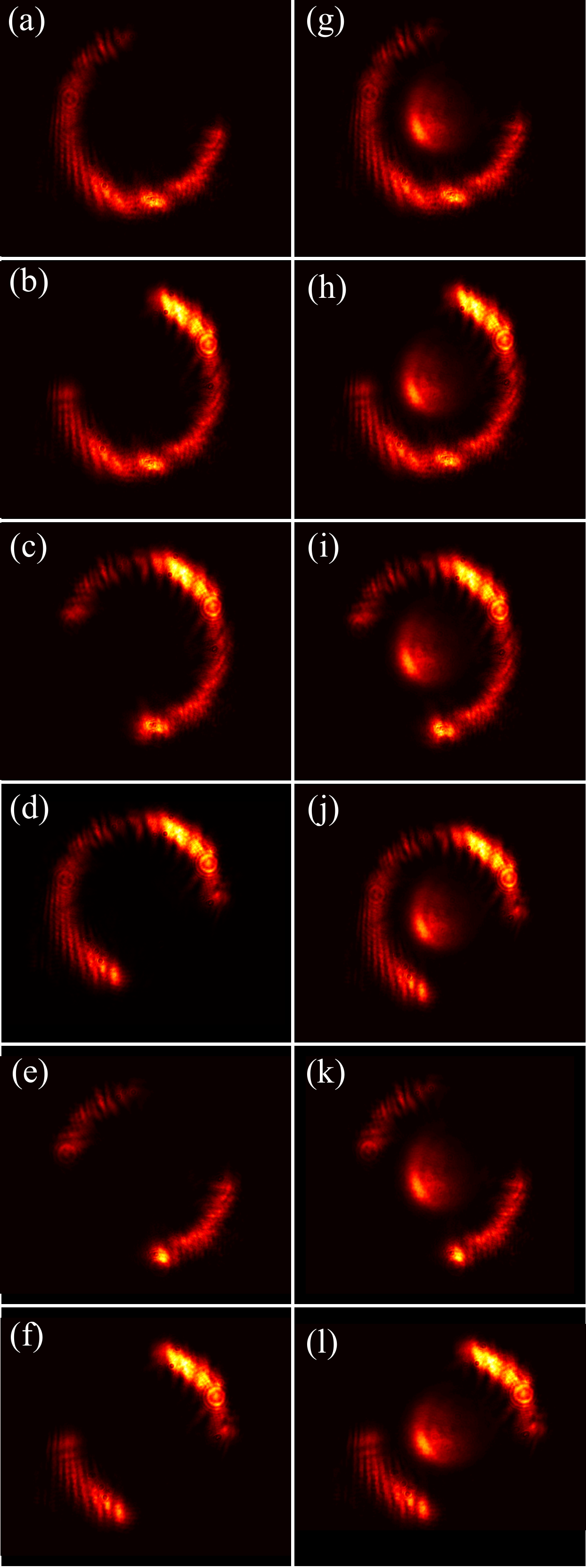}
\caption{The results of the Segmented space division multiplexing experiment. (a)-(f): The multiplexed LG beams. Each of these beams show a specific arrangement of bits. (g)-(l): Multiplexed beams coupled with the filament. }
\label{fig:LG_modulation}
\end{figure}\\
The resulting beam has 16 possible combinations that are realized by individually modulating the intensity of each segment. Therefore, no signal is state 0000 and the full donut beam is the 1111 state as shown in Figure \ref{fig:bits}.\\
This results in a noise resilient FSO scheme, with a signal to noise ratio (SNR) of 17.8 dB. The SNR was calculated using: 
\begin{equation}
\rm{SNR} = 20\times\log_{10}\left(\frac{Signal}{Noise}\right)
\label{eq:SNR}
\end{equation} 
The signal and the noise are characterized using a power-meter set up at the place of the sCMOS camera. The power meter is placed at the detection zones as indicated in Figure \ref{fig:LG_nomod}--(d).
The signal values are measured with parts of the LG beam illuminating the power-meter together with the ultrafast signal. To calibrate the noise, the LG beam is blocked and the pump is left on. For each sets of measurements, the power from each signal is kept constant.\\
Various examples of the modulated beam can be seen in Figure \ref{fig:LG_modulation}. The left column Figure  \ref{fig:LG_modulation}$-$(a) through (f) shows just the modulated LG beam, whereas the right column Figure \ref{fig:LG_modulation}$-$(g) through (l) shows the corresponding coupled beams. It can be seen that this scheme produces a 4$-$bit multiplexed signal with minimal crosstalk between bits and that the signal is unaffected by the presence of the filament. In Figure \ref{fig:LG_modulation} six combinations are represented  as a showcase rather than the full list of all the 16 combinations for the 4-bit system. Furthermore, we demonstrate that each quadrant can be modulated independently.
\subsection{\small\rm\bf Gaussian array}
\begin{figure}[htbp]
\centering
\includegraphics[width=0.95\linewidth]{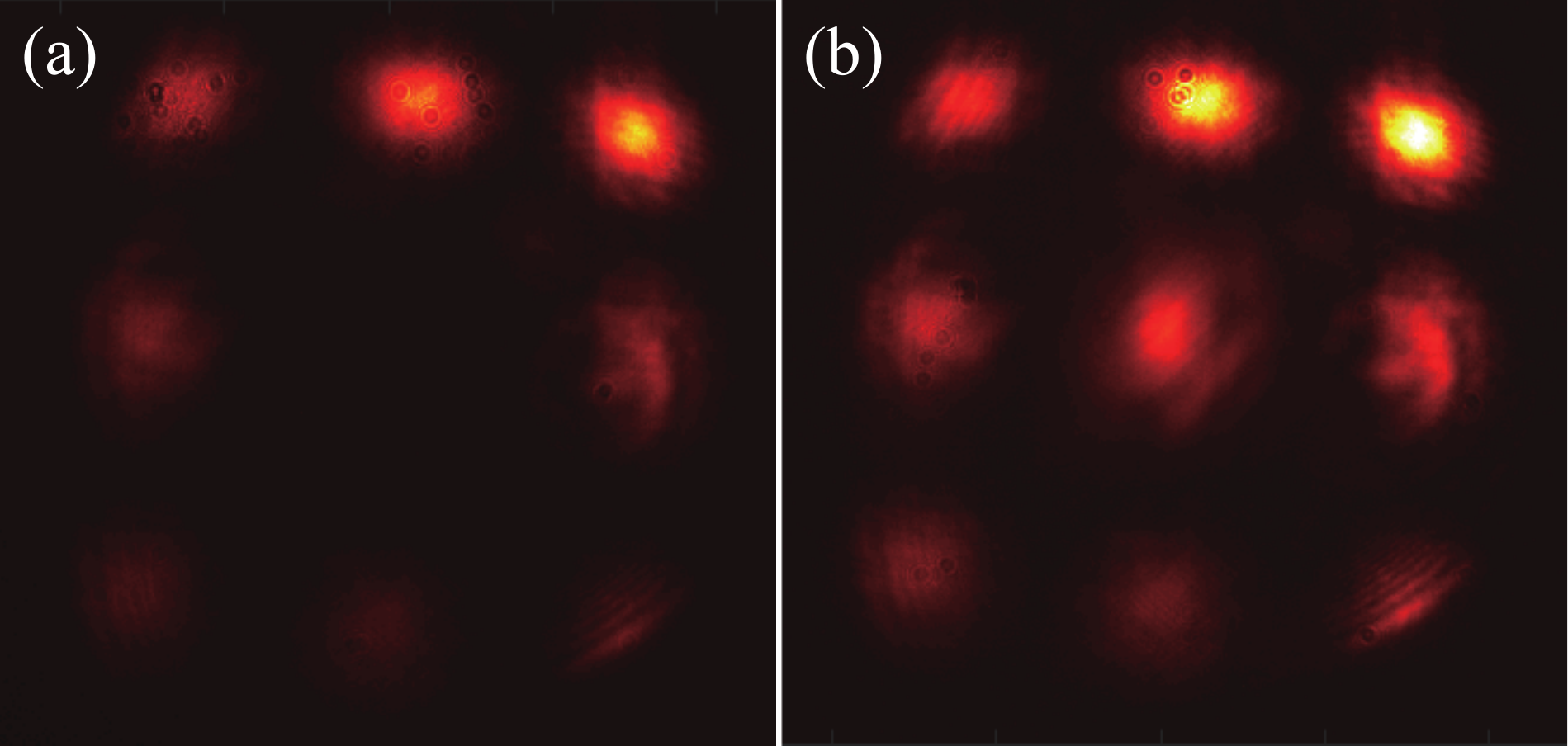}
\caption{(a): 8 Gaussian beam array generated using the SLM. (b): 8 Gaussian beam array coupled with the filament.}
\label{fig:GaussianArray}
\end{figure}

Another structured light beam that can be coupled around the filament is the Gaussian array.  As shown in Figure \ref{fig:GaussianArray}$-$(a), 8 separated Gaussian beams are arranged in a square, the center of which is left empty by design. The Gaussian array can be arranged to form numerous shapes according to the configuration of the phase mask.
\\
Thus the Gaussian array provides a unique solution to create an information channel that can be configured to closely match the spatial characteristics of the quasi-transparent channel. Like LG beams, the Gaussian array can also be used in a SSDM scheme. Each beam representing a bit can be individually controlled via the SLM. In this way the Gaussian array can form an $n$--bit SSDM communications system. Scalability of this system expands the potential of FSO systems to work at much higher capacities than current solutions. However, the Gaussian beams do not possess the non-diffractive characteristics of the LG beam. Therefore, the probability of crosstalk between bits is increased. 
\subsection{\small\rm\bf Bessel-Gaussian beams and interaction with filament}
\begin{figure}[h]
\centering
\includegraphics[width=0.95\linewidth]{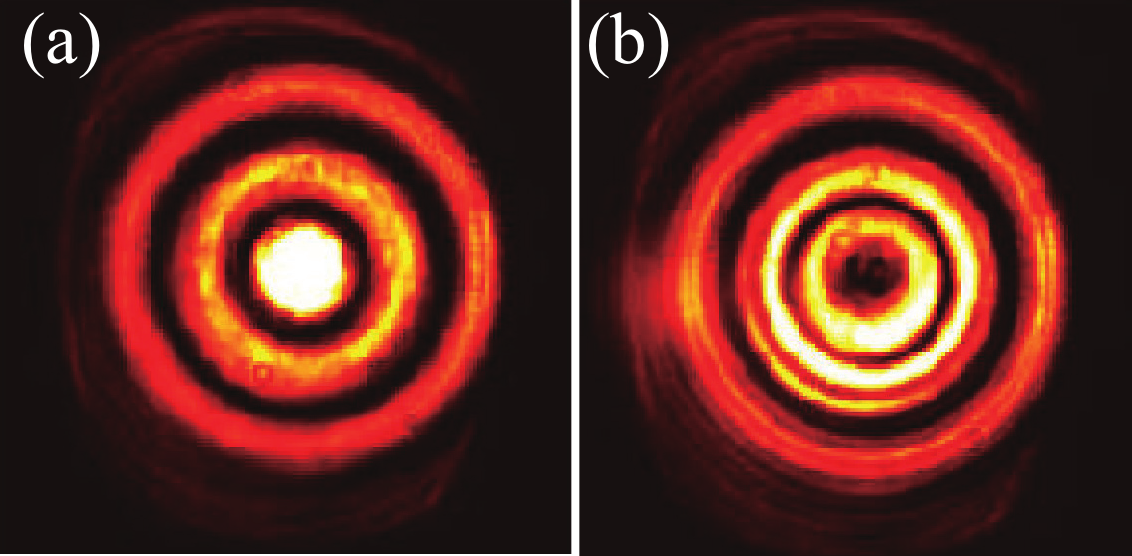}
\caption{(a): 0$^{th}$ order Bessel-Gaussian (BG) beam with the femtosecond beam blocked. (b): BG coupled with the filament. The filament causes the light to diverge from the center and hence we see a hole despite it being a 0$^{th}$ order BG.}
\label{fig:Bessel_interaction}
\end{figure}

In order to demonstrate how the filament interacts with other light sources, we replaced the SLM in Figure  \ref{fig:setup} with an axicon ($\alpha$ = 0.5$^{\circ}$) and the MC with a DM. The resulting 0$^{th}$ order Bessel-Gaussian (BG) beam was expanded and was coupled with the filament. Figure \ref{fig:Bessel_interaction}$-$(a) shows the BG beam while the femtosecond beam is blocked. In figure \ref{fig:Bessel_interaction}.(b) the femtosecond beam is unblocked and forms the filament through the Gaussian center of the BG beam. As shown in the figure, the filament causes the center of the BG beam to diverge outwards.
This interaction is not contained to the center of the BG beam, with the effects visible in the 1$^{st}$ and to a lesser degree in the 2$^{nd}$ concentric rings. This interaction introduces crosstalk between channels and is detrimental to the overall performance of the system. Taking Figure \ref{fig:Bessel_interaction} as an example, the SNR drops from 15.6 dB in Figure \ref{fig:Bessel_interaction}$-$(a) to 9.4 dB in Figure \ref{fig:Bessel_interaction}$-$(b). Research conducted previously demonstrated the self-reconstruction properties of BG beams \cite{Zhao:2019}. This is particularly useful when it comes to FSO as, partially obstructing the beam will lead to self-reconstruction. However this is only possible with higher order BG beams as the 0$^{th}$ order BG beam will inevitably be affected by the filament.
\section{\rm Conclusion}
We have experimentally demonstrated a novel and robust scheme for FSO by embedding a femtosecond laser filament within non-diffractive structured light beams with donut shaped intensity distribution. We have devised a new method to couple the signal carrying the information to the filament using an in-house mirror coupler. We have also proposed a spaced division multiplexing scheme that showed minimal crosstalk between channel while significantly increasing the capacity of the data transmitted. 
We have coupled the filament with beams possessing Gaussian-like distribution to show that the implementation of this method will lead to crosstalk and interference with the filament. We believe that the hybrid method of communication we have presented will result in robust FSO communication systems, which can be implemented to compensate for losses induced by difficult weather conditions.\\
\\
{\bf Funding:} {\small This work is supported in part by the National Geospatial Intelligent Agency grant $\#$HM04762010012.}\\
\\
\smallskip
{\bf Disclosures:}\textmd{ \small The authors declare no conflicts of interest.}

\printcredits


\end{document}